\documentclass[aps,pra,floatfix,twocolumn,tightenlines,superscriptaddress,amsmath,amssymb]{revtex4}

\usepackage{graphicx}
\usepackage{amssymb}
\usepackage{color}
\usepackage{psfrag}
\usepackage{ifsym}
\usepackage{epstopdf}

\newcommand{\ket}[2] {| #1 \rangle_{#2}}

\newcommand{\ii} {\textbf{i}}

\begin{document}

\title{Engineered optical nonlinearity for quantum light sources}

\author{Agata M.~Bra{\'n}czyk}
\affiliation{Center for Quantum Computer Technology, Department of Physics, The University of Queensland, QLD 4072, Australia}

\author{Alessandro Fedrizzi} 
\affiliation{Center for Quantum Computer Technology, Department of Physics, The University of Queensland, QLD 4072, Australia}

\author{Thomas M. Stace}
\affiliation{Department of Physics, The University of Queensland, QLD 4072, Australia}

\author{Tim C. Ralph} 
\affiliation{Center for Quantum Computer Technology, Department of Physics, The University of Queensland, QLD 4072, Australia}

\author{Andrew G. White}
\affiliation{Center for Quantum Computer Technology, Department of Physics, The University of Queensland, QLD 4072, Australia}

\begin{abstract}  
Many applications in optical quantum information processing benefit from careful spectral shaping of single-photon wave-packets. In this paper we tailor the joint spectral wave-function of photons created in parametric downconversion by engineering the nonlinearity profile of a poled crystal. We design a crystal with an approximately Gaussian nonlinearity profile and confirm successful wave-packet shaping by two-photon interference experiments. We numerically show how our method can be applied for attaining one of the currently most important goals of single-photon quantum optics, the creation of pure single photons without spectral correlations.
\end{abstract}

\maketitle

Spontaneous parametric downconversion is a nonlinear optical process in which a photon from a pump laser, incident on a nonlinear birefringent crystal, converts into two single photons under conservation of energy and momentum. Photon sources based on this phenomenon are an ubiquitous tool for quantum computation \cite{Kok2007}, quantum communication \cite{gisin2007qc} and quantum metrology \cite{higgins2007ehp, nagata2007bsq}. They are also becoming increasingly important in more specialised applications such as quantum imaging \cite{brida2010ers}, quantum lithography \cite{boto2000qio} or optical coherence tomography \cite{nasr2008ubg}. As these experiments evolve, more stringent requirements are placed on the characteristics of the quantum state of emitted light. In particular, to produce high-purity heralded, or even near deterministic single photons, the spectral shape and correlations of the created photon pairs must be carefully engineered. 

The most common method for spectral engineering is filtering, however, this can lead to loss and mixing. A more sophisticated method involves shaping the spectrum at the source. Consider downconversion in a quasi-phasematched (QPM) crystal with a poling period $\Lambda$ \cite{Fejer1992}, where the crystal domain is inverted whenever the pump and downconversion fields acquire a phase mismatch $\Delta k{=}2 m\pi/\Lambda$---where $m$ is an odd integer---allowing phase-matching of a wide range of wavelengths in different nonlinear materials. There are several methods to control the joint spectral amplitudes of photons created in downconversion in a QPM crystal \cite{Fejer1992,imeshev2001psd}. For example, imposing a linear chirp on the poling period $\Lambda$ has been used for the generation of ultra-broad-spectrum, top-hat shaped photons \cite{nasr2008ubg} for optical coherence tomography. However, the currently known methods involve changing $\Lambda$, which may be incompatible with stringent phasematching conditions.

In this paper, we consider type-II downconversion in a QPM crystal with a longitudinally non-uniform grating. We can synthesise photon pairs with arbitrary spectral amplitudes by modulating the nonlinearity profile $\chi(z)$ of a crystal through different-order poling \textit{without} changing the phase-matching conditions. We tailor a spectral photon-pair amplitude with an approximately Gaussian profile, which is critical for high purity photon generation, as we will discuss later. In addition, Gaussian spectra are known to be optimal for temporal mode matching \cite{rohde2005opq}---a critical consideration in any experiment involving single photons.

Theoretically, the two-photon component of the optical state is described by \cite{rubin1994ttp}
\begin{eqnarray}\label{eq:two_photon_state}
\ket{\psi}{}=\int\int d\omega_i d\omega_sf(\omega_i,\omega_s)\hat{a}_i^{\dagger}(\omega_i)\hat{a}_s^{\dagger}(\omega_s)\ket{0},
\label{eq:spdc-state}
\end{eqnarray}
where $f(\omega_i,\omega_s)=\alpha(\omega_i+\omega_s)\Phi(\omega_{i},\omega_{s})$ is the joint spectral amplitude of the created photons in the \textit{idler} and \textit{signal} modes respectively (for details, refer to Appendix \ref{sec:hamiltonian}). The spectral properties of downconverted photons can be manipulated via the pump envelope function $\alpha(\omega_i+\omega_s)$ \cite{valencia2007swe}, or as we show here, the phase matching function (PMF) $\Phi(\omega_{i},\omega_{s})$. We use a monochromatic pump $\alpha(\omega_i+\omega_s)=\delta(\omega_i+\omega_s-\mu_p)$ where $\mu_p$ is the pump frequency, and directly tailor 
\begin{eqnarray}\label{eq:phi}
\Phi(\omega_{i},\omega_{s})=\sqrt{2\pi}\int_{-\infty}^{\infty}\chi(z)e^{{-}\ii\Delta k(\omega_{i},\omega_{s}) z}dz,
\label{eq:pmf}
\end{eqnarray}
which is the Fourier transform of $\chi(z)$ and $\Delta k$ is the phase mismatch. 

According to equation \ref{eq:phi}, the phase matching function of a standard crystal with a uniform nonlinearity profile is $\Phi(\omega_i,\omega_s)=\mathrm{sinc}(\Delta kL/2)$. However, to generate a Gaussian phase matching function, we require a crystal with a Gaussian nonlinearity. While it is non-trivial to directly change the material properties, we can make use of higher-order poling to realise a variety of nonlinearity strengths. In a poled structure, the \emph{effective} nonlinearity scales with the poling order $m$ as $\chi_{\textrm{eff}}{=}2\chi/\pi m$. For odd $m$, $m$th order QPM can be achieved by reversing the direction of the poling every $m$ coherence lengths, defined as $L_c=\Lambda/2$. Even-order QPM can be achieved by combining two odd orders. 

 \begin{figure*}[t!]
  \begin{center}
 \includegraphics[width=1\textwidth]{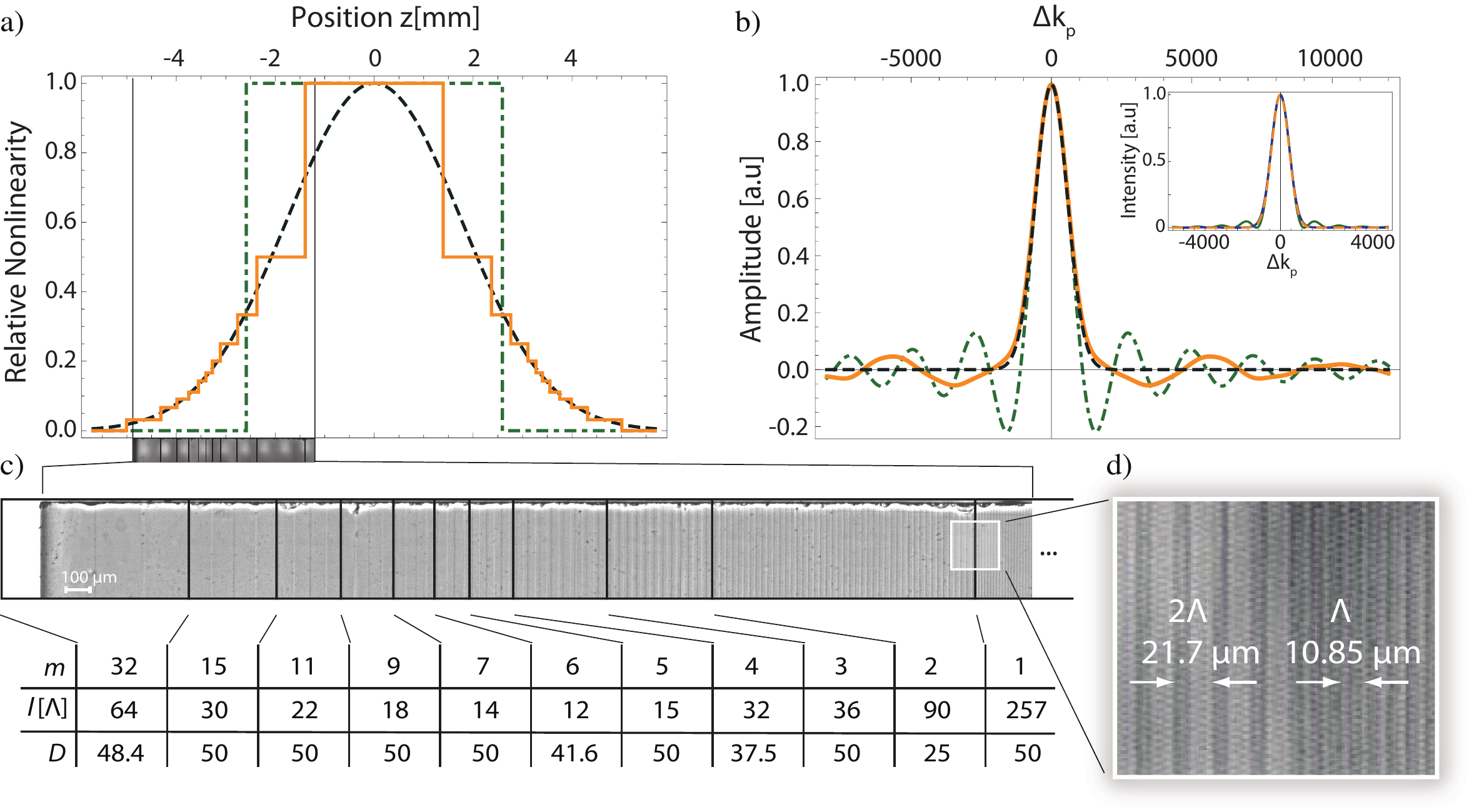}
  \end{center}
    \caption{a) Nonlinearity profile for the cpKTP crystal $\chi_{\textsc{t}}(z)$ (orange line) and target Gaussian profile $\chi_{\textsc{g}}(z)=\exp(-(z/L_\textrm{eff})^2/\gamma)$ (black dashed line) with effective length $L_\textrm{eff}=5.67 mm$ (green dot-dashed line) and $\gamma\approx0.193$ (see Appendix \ref{sec:Gauss_approx}). b) Phase-matching function amplitudes and intensities (inset) for the cpKTP (orange line) compared to a ppKTP of the same \textit{effective} length $L_\textrm{eff}$ (green dot-dashed line) and target Gaussian profile $\Phi_{\mathrm{G}}(\omega_i,\omega_s)=\exp(-\gamma (\Delta kL/2)^2)$ (black dashed line). c) Magnified image of part of the custom-poled KTP crystal. Vertical lines separate sections with constant effective nonlinearity, with their poling order $m$, length $L$ and poling duty cycle $D$. d) Magnified view of the transition from poling order $m{=}1$ to $m{=}2$ \cite{Raicol}. Due to a slight mismatch between design and actual domain lengths, the crystal was shortened by a few tens of $\mu m$ on one side.} 
  \label{fig:prob2ext}
\end{figure*}

We exploited this feature to design a crystal consisting of a number of discrete sections, each with a different $\chi_{\textrm{eff}}$, discretely approximating the desired Gaussian shape, and custom-poled a 10 mm long Potassium Titanyl Phosphate (cpKTP) crystal accordingly.  Details on the actual design of our tailored crystal can be found in Appendix \ref{sec:tailoring}. Our technique decreases the overall nonlinearity and reduces the \textit{effective} length of the structure, Appendix \ref{sec:Gauss_approx}. We therefore expect less photon-pair yield and broader bandwidths when compared to a standard periodically poled KTP (ppKTP) with the same length and phase-matching. 

Fig. \ref{fig:prob2ext}a) shows the tailored nonlinearity profile $\chi_{\textsc{t}}(z)$, defined in Eq. (\ref{eq:chi_m}) in Appendix \ref{sec:tailoring}, together with the target Gaussian profile $\chi_{\textsc{g}}(z)$. The corresponding PMF, obtained from the inverse Fourier transform of $\chi_{\textsc{t}}(z)$, is very similar to a Gaussian function, as shown in Fig.~\ref{fig:prob2ext}b). Compared to the sinc-shaped phase matching function of a ppKTP of the same \textit{effective} length as the cpKTP (5.67 mm), the side lobes on either side of the central peak are significantly suppressed. This becomes even more evident when considering the spectral intensity (see inset). 

A microscopic image of part of the custom-poled KTP (cpKTP) crystal is shown in Fig.~\ref{fig:prob2ext}c). One can clearly see the individual sections with different poling orders, which line up with the theoretical design almost perfectly. Fig.~\ref{fig:prob2ext}d) shows a magnified view of a transition between poling-order sections $m{=}1$ and $m{=}2$.

\begin{figure*}[t!]
  \begin{center}
 \includegraphics[width=1\textwidth]{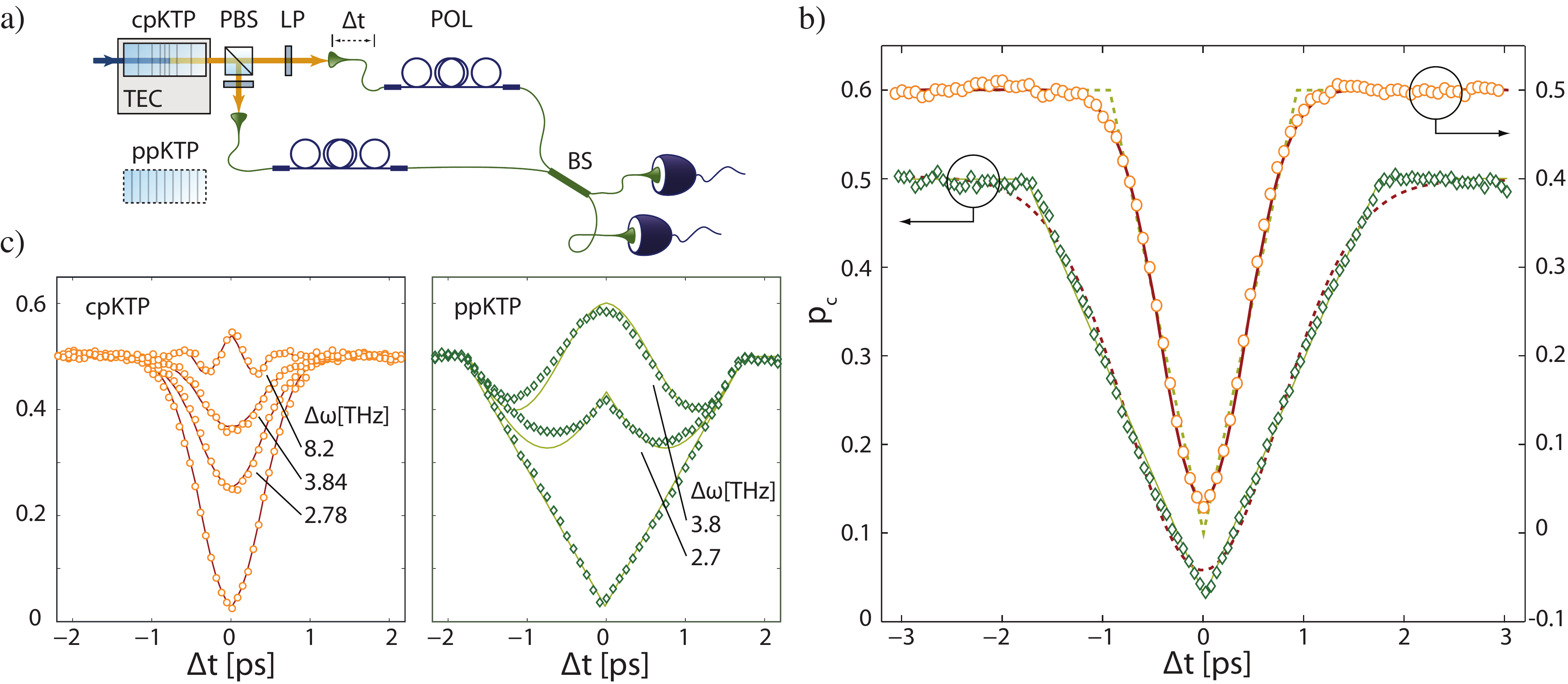}
  \end{center}
    \caption{a) Experimental scheme. The crystals (cpKTP, ppKTP) were temperature-stabilised (TEC) and pumped by a $410$~nm, grating-stabilised diode laser. The emitted orthogonally polarised photon pairs were split at a polarising beamsplitter (PBS) and coupled into single-mode fibres equipped with polarisation-controllers (POL). They were then superposed at a 50/50 fiber beamsplitter (BS) and detected in coincidence. We obtained two-photon interference patterns by changing the delay $\Delta t$ with a motorised translation stage. The only filters in use were two RG715 long-pass filters (LP).
b) Two-photon interference patterns for the cpKTP (red circles) compared to a standard ppKTP (green diamonds). The solid lines show the theoretical values, calculated from the respective PMF for each crystal. The reduced chi-square values of these fits are $3.07$ and $5.51$, respectively. The dashed lines show least-square fits of a triangular pattern to the tailored crystal data and a Gaussian fitted to the normal crystal, with reduced chi-square values of $50.59$ and $23.10$, underlining the strong divergence from these shapes.    
c) Spatial quantum beating for various center-frequency detunings $\Delta\omega=\omega_{i}-\omega_{s}$. The lines show the ideal values, calculated from the respective PMF. All probabilities $p_{c}$ for b) and c) were obtained by normalising detected pairs to twice the averaged counts outside the coherence length. The only free parameter for theory values was the interference visibility of $\sim95\%$. All error bars are smaller than symbol size.} 
  \label{fig:experiment}
\end{figure*}

We tested our custom-poled crystal in a typical downconversion setup, see Fig.~\ref{fig:experiment}a), comparing it to a $10$~mm long ppKTP crystal ($\Lambda{=}10.95~\mu$m). Due to the reduced overall effective nonlinearity, we expect a relative photon pair rate of $34.4\%$. The measured rate (detected without a beam-splitter) was ${\sim}10$~kpairs/s for the custom-poled and ${\sim}33$~kpairs/s for the standard crystal, respectively. This corresponds to a relative yield of $\sim30.4\%$; we attribute the small reduction in efficiency to the fact that the custom-poled crystal, in contrast to our standard crystal, was not anti-reflection coated.

We verified the spectral amplitude of bi-photons created in the cpKTP in a two-photon interference experiment. When two indistinguishable photons mix on a symmetric beamsplitter, they will always be found in the same output port. This phenomenon was first reported in the landmark experiment by Hong, Ou and Mandel \cite{hong1987mst}, who observed a dip in the photon-pair detection probability as a function of the temporal delay between the input photons. Theoretically, the shape of this coincidence dip assumes the inverted Fourier transform of the absolute square of the PMF, which can be readily calculated from the two-photon state in Eq. (\ref{eq:spdc-state}), see \cite{hong1987mst}.

The recorded interference patterns for the cpKTP, and the standard ppKTP are shown in Fig. \ref{fig:experiment}b), along with the theoretical interference patterns calculated straight from the PMFs in Fig.~\ref{fig:prob2ext}b). The bandwidth difference results from the different \textit{effective} lengths of the two crystals.  The interference pattern for the standard crystal is triangular, just as expected for the sinc-shaped PMF \cite{fedrizzi2009arh}. The pattern for the custom-poled crystal departs from the triangular shape and indeed approximates a Gaussian. The interference visibility was $\sim95\%$ for both crystals, confirming that the indistinguishability of the downconverted photons was not compromised by the crystal modulation. 

To  further explore the underlying spectral correlations in the PMF, we measured spatial quantum beating patterns. We detuned the center frequencies of the downconversion photons via a  change in crystal temperature away from its optimal value for collinear, degenerate quasi-phasematching and again observed two-photon interference \cite{fedrizzi2009arh}. The results in Fig.~\ref{fig:experiment}c) show that the custom-poled crystal exhibits less distinct beating, in particular, less anti-bunching, i.e. coincidence probability values above the random level of $0.5$. The maximum value for the cpKTP was $0.546\pm0.005$ compared to $0.586\pm0.003$ for the standard crystal, a significant reduction relative to the base-line of $0.5$. The theory values for the ppKTP were adopted from \cite{fedrizzi2009arh} directly. For the cpKTP we repeated the calculation in \cite{fedrizzi2009arh} using the respective custom PMF shown Fig. \ref{fig:prob2ext}b).

The observed anti-bunching occurs when the frequency-detuned spectral wavefunction (i.e. the joint spectral amplitude) of the two-photon state is partially anti-symmetric, which in turn reveals the frequency entanglement intrinsic to downconversion \cite{wang2006qtt,fedrizzi2009arh,eckstein2008bfm}. A Gaussian spectral amplitude is always positive and therefore does not have anti-symmetric components, which explains the significantly reduced beating in the interference patterns of the custom-poled crystal. All measured interference patterns for our custom crystal agree exceedingly well with the patterns calculated from the theoretic PMF. 

One situation in which this method will be useful is the generation of pure heralded single photons. Even in a group-velocity-matched configuration with a symmetric joint spectral amplitude, the maximum purity of the heralded single photon state is, due to the PMF sinc profile \cite{Branczyk2009}, limited to $0.81$ (for ppKTP with a $788$ nm pump \cite{Grice2001,Kim2002,URen2005} without spectral filtering). This purity affects the two-photon interference visibility between heralded single photons and thus ultimately the quality of optical quantum gates or multi-photon states generated in post-selection. We numerically compare two designs for cpKTP crystals (see Appendix \ref{sec:crystal_design}), characterised by their maximum poling order ($m{=}1$ and $m{=}2$), with a standard ppKTP crystal following \cite{Branczyk2009}. Table \ref{tab:compare crystals} shows the calculated purities for our two designs compared with a standard crystal. Taking $m_{\mathrm{min}}{=}1$ yields a coarse approximation to the Gaussian function, but the purity $P$ improves substantially to $0.97$. A better approximation to the Gaussian is achieved if $m_{\mathrm{min}}{=}2$, then $P{=}0.99$. For further detail, refer to Appendix \ref{sec:crystal_design}.
\begin{table}[h]
\caption{Numerical comparison of the purity $P$ of heralded single-photons of a standard crystal and two cpKTP crystals of length $L$, in a group-velocity-matched scenario. The effective length of both cpKTP crystals is $L_\textrm{eff}{=}24.2$mm.}
\vspace{4mm}
\centering
\begin{tabular}{c c  c  c c}
  \hline
Crystal & $L$ [mm] & Purity $P$  \\
  \hline
ppKTP & 24.2  & 0.81     \\
  cpKTP $m_{\mathrm{min}}=1$ & 40.5 & 0.97     \\ 
  cpKTP $m_{\mathrm{min}}=2$ & 41.6 & 0.99    \\ 
  \hline
\end{tabular}
     
  \label{tab:compare crystals}
\end{table}

In conclusion, we demonstrated longitudinal shaping of single-photon wave-packets via indirect modulation of the nonlinearity of a crystal. Our method can be used to generate other spectral profiles of interest, such as a triangle or a top-hat, see Appendix \ref{sec:other_PMF}. A comb-like nonlinearity structure, for example, would allow the direct, lossless generation of frequency-bin qubits \cite{ramelow2009dtc,olislager2010fbe}. In addition, the custom-poled crystals can be used in single-crystal sources for polarization entanglement, such as the Sagnac-type confgurations reported in \cite{kim2006pss,fedrizzi2007wtf}, or even in a single-pass scheme, combined with the related technique of using two interlaced first-order poling periods for concurrent type-II downconversion of frequency non-degenerate photons \cite{thyagarajan2009gpe}. Furthermore, we expect this technique to have applications in classical nonlinear optics, e.g., in second harmonic generation, similar to the spectral shaping techniques previously demonstrated for this regime \cite{Fejer1992,imeshev2001psd}.

Since our technique allows manipulation of the phasematching profile, it complements group-velocity matching, which is commonly achieved by controlling the orientation and width of the phasematching function. A Gaussian shape vastly improves the purity of heralded photons from SPDC. The inevitable reduction in the effective nonlinearity is an acceptable tradeoff given that modern crystals have drastically reduced pump power requirements. In addition, compared to the alternative of spectral filtering, one can actually pump at a much higher power without introducing photon number mixedness. This allows the creation of purer multi-photon states for quantum information processing, e.g. Fock states with high photon number \cite{Branczyk2009}.

It will be worthwhile to consider nonlinearity engineering for four-wave-mixing photon-pair sources in photonic-crystal fibres, where the sinc-shaped phase-matching function has been identified as a major problem \cite{halder2009nti}. However, group-velocity matching in these materials is already a non-trivial task which will inevitably be further complicated by modulation of the non-linearity.

\section*{Acknowledgments}
We thank B.~Kuhlmey for helpful discussions and T.~McRae for help with imaging. We acknowledge support by the ARC Discovery and Federation Fellow programs and an IARPA-funded ARO contract.

\appendix
\section*{Appendix}

\section{Type II SPDC Hamiltonian}\label{sec:hamiltonian}
For type-II up- or down-conversion, the evolution inside the crystal is governed by the operator $U(t)=\mathcal{T}\mathrm{exp}({-}\frac{\imath}{\hbar}\int _{t_0}^{t'} dtH(t))$ where $\mathcal{T}$ is the time-ordering operator and
\begin{eqnarray}\nonumber
 H(t)&=&A\int d\omega_i  d\omega_s d\omega_p \alpha(\omega_p) e^{\ii\Delta \omega t}\hat{a}_i^{\dagger}(\omega_i)\hat{a}_s^{\dagger}(\omega_s)\\
 &&\times\int_{{-}L/2}^{L/2}dz\chi^{(2)} e^{{-}\ii\Delta k(\omega_i,\omega_s,\omega_p) z}+\mathrm{H.c.}\,,
\end{eqnarray}
is the multimode Hamiltonian \cite{Grice1997,Branczyk2009}, where $\Delta \omega = \omega_i+\omega_s-\omega_p$, $L$ is the length of the crystal, $A$  is a constant proportional to the nonlinearity and $\Delta k(\omega_{i},\omega_{s},\omega_{p})=k_{p}(\omega_{p})-k_{i}(\omega_{i})-k_{s}(\omega_{s})$ is the phase mismatch. Typically, the nonlinearity does not vary over the length of the crystal and we can rewrite the spatial integral $\int_{{-}L/2}^{L/2}dz\chi^{(2)} e^{{-}\ii\Delta k(\omega_i,\omega_s) z}$ as the Fourier transform (FT) of a rectangular function $\sqrt{2\pi}\int_{-\infty}^{\infty}\chi_{\mathrm{r}}(z)e^{{-}\ii\Delta k(\omega_i,\omega_s) z}dz$, where $\chi_{\mathrm{r}}(z)=\chi^{(2)}(u(z+L/2)-u(z-L/2))$ and $u$ is the Heaviside step function. Evaluating the spatial integral yields the result $L\mathrm{sinc}(\Delta k(\omega_i,\omega_s)L/2)$ where $\mathrm{sinc}(x)=\sin(x)/x$, giving rise to the following form for the Hamiltonian 
\begin{eqnarray}\nonumber
H(t)&=& AL\int d\omega_i d\omega_sd\omega_p \alpha(\omega_p)\Phi(\Delta k(\omega_i,\omega_s,\omega_s))\\\label{eq:ham}
&&\times e^{\ii\Delta \omega t}\hat{a}_i^{\dagger}(\omega_i)\hat{a}_s^{\dagger}(\omega_s)+\mathrm{H.c.}\,,
\end{eqnarray}
where 
\begin{eqnarray}
\Phi(\Delta k(\omega_i,\omega_s,\omega_p))=\mathrm{sinc}\Big(\frac{1}{2} \Delta k(\omega_i,\omega_s,\omega_p)L\Big)\label{eq:PMF}
\end{eqnarray}
 is the phase-matching function. Note that by picking the spatial integration to be centered around $z=0$, it is possible to eliminate a phase term which would normally be present in equation (\ref{eq:ham}). 
 
For a periodically poled crystal the phase mismatch,  $\Delta k_p(\omega_{i},\omega_{s},\omega_p)=\Delta k(\omega_{i},\omega_{s},\omega_{p})-2\pi/\Lambda$, includes the effect of the periodic domain inversion in QPM. 
The purpose of this domain inversion is to undo the $m\pi$---where $m$ is an odd integer---phase error accumulated by the pump and downconverted fields, by introducing an additional $\pi$ phase shift to $\Delta k$ at the point of inversion. For successful QPM, the poling period therefore has to fulfil the condition $\Lambda=m 2\pi/\Delta k$, where $m$ is the QPM order.

Taking the first order term of the Taylor series expansion of $U(t)\ket{0}{}$ and evaluating the time and pump frequency integrals yields the two-photon state
\begin{eqnarray}
\ket{\psi}{}=\int\int d\omega_i d\omega_sf(\omega_i,\omega_s)\hat{a}_i^{\dagger}(\omega_i)\hat{a}_s^{\dagger}(\omega_s)\ket{0}{}\,,
\end{eqnarray}
where $f(\omega_i,\omega_s)=\alpha(\omega_i+\omega_s)\mathrm{sinc}(\Delta k(\omega_i,\omega_s)L/2)$ and  $\Delta k(\omega_{i},\omega_{s})=k_{p}(\omega_{i}+\omega_{s})-k_{i}(\omega_{i})-k_{s}(\omega_{s})$.
 
\section{Tailoring the crystal nonlinearity} \label{sec:tailoring}
To tailor the crystal nonlinearity, we treat each crystal section $s$ as a rectangular function with a nonlinearity inversely proportional to the poling order $m_s$. The nonlinearity profile for the custom-poled crystal is then given by
\begin{equation}\label{eq:chi_m}
\chi_{\textsc{t}}(z)=\sum_{s=1}^{21}\frac{1}{m_s}u\Big(\frac{1}{2}\sum_{r=1}^{s}m_rn_r\Lambda-z\Big)
\times u\Big(z-\frac{1}{2}\sum_{r=1}^{s-1}m_rn_r\Lambda\Big),
\end{equation}
where $u$ is the Heaviside step function, $m_r$ is the poling order of the $r$th section, $n_r$ is the number of domains within the $r$th section and $\Lambda{=}10.85~\mu$m, for type-II, first-order QPM of $410$~nm${\rightarrow}820$~nm${+}820$~nm.  The profile $\chi_{\textsc{t}}(z)$ is plotted in Fig. \ref{fig:prob2ext}a).

The design of the nonlinearity profile is subject to two constraints. First, the nonlinearity of each section is limited to discrete values proportional to $1/m$. Larger values of $m$ provide smoother transitions between successive nonlinearites, however this leads to a greatly reduced photon creation rate. Second, the width of each section must be an integer number of $m\Lambda/2$ and a minimum of $2m\Lambda$, therefore, larger values of $m$ may demand prohibitively long sections. The ratio between positively and negatively poled regions---known as the duty cycle $D=l/m\Lambda$ where $l$ is the length over which the sign of the nonlinear coefficient remains constant---was chosen to be $50\%$ for odd values of $m$ and as close as possible to $50\%$ for even values, as is shown in Fig. \ref{fig:experiment}c). 

While the model presented here is not strictly valid due to the small number of domains within each section of $\chi_{\textsc{t}}(z)$, modeling at the domain level, Appendix \ref{sec:detailed_model}, shows good agreement with the basic model in the frequency range over which the detectors are sensitive. The edges of the spectral response function of the detectors used in this experiment lie in the region of negligible amplitude and, as opposed to spectral filters, will not give rise to the photon-number mixing described in \cite{Branczyk2009}. 
 
\section{Gaussian approximation to sinc function}\label{sec:Gauss_approx}
To determine the exact shape of the target Gaussian function for the nonlinearity profile, we match the width of the desired Gaussian PMF with the sinc PMF of the form, $\mathrm{sinc}(\Delta kL_\textrm{eff}/2)$, that would be generated by a standard crystal. The appropriate function is $\Phi_{\mathrm{G}}(\omega_i,\omega_s)=\exp(-\gamma (\Delta kL_\textrm{eff}/2)^2)$ where the parameter $\gamma\approx0.193$ is derived from matching the FWHM of the two functions. We refer to $L_\textrm{eff}$ as the \textit{effective} length, as it does not correspond to the actual length of the final Gaussian shaped crystal, but rather the length of the hypothetical standard crystal.

  \begin{figure}[b!]
  \begin{center}
 \includegraphics[width=8cm]{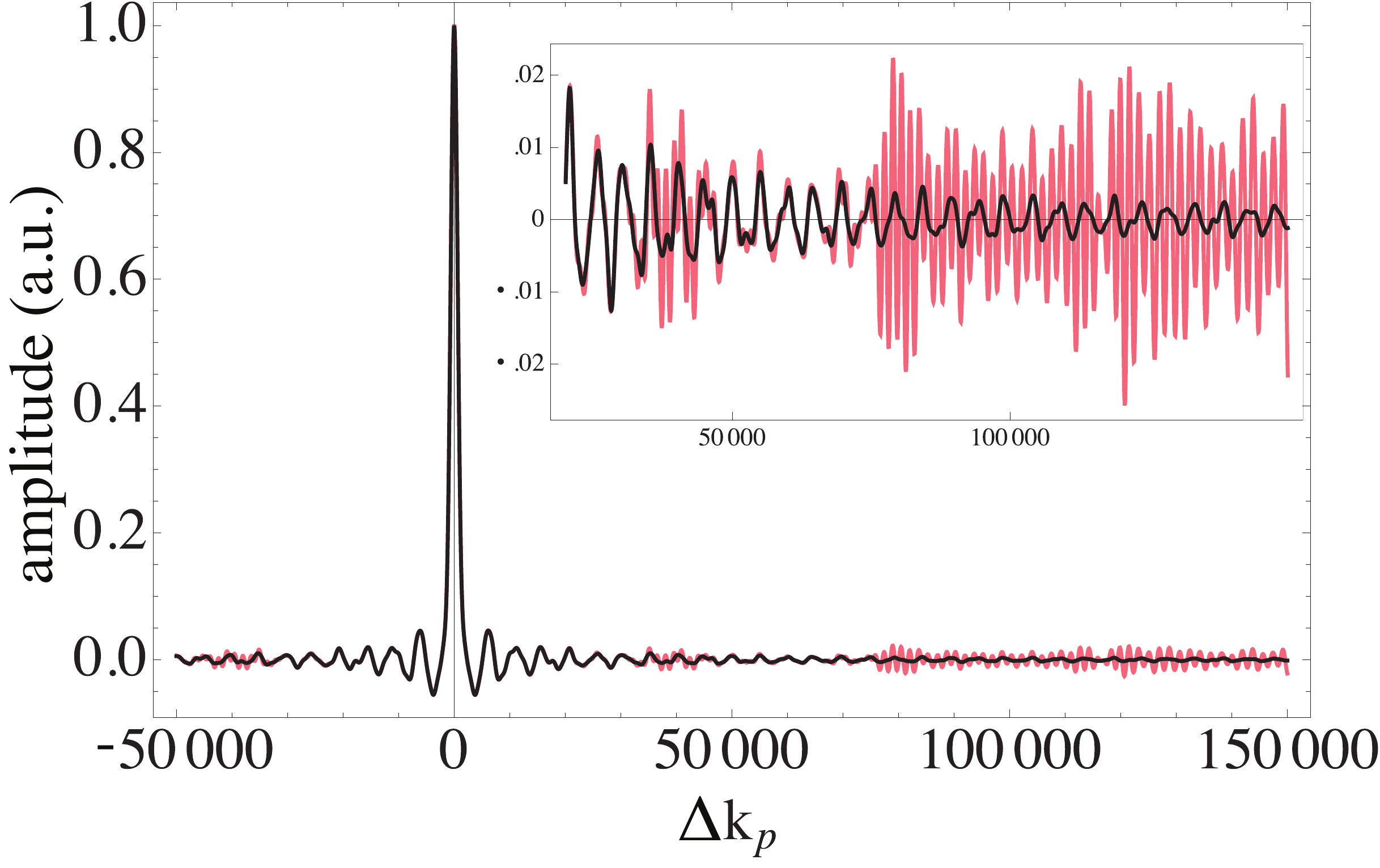}
    \vspace{-.5cm}
  \end{center}
    \caption{Phase matching functions generated from the basic model (black solid line) and the detailed model (light red line). The inset shows a magnified portion of the PMFs, detailing the deviation between the models. } 
  \label{fig:comp_zoom}
\end{figure}
 
\section{Detailed Model}\label{sec:detailed_model}
We modeled each section of the crystal as having a nonlinearity inversely proportional to the poling order $m$. This approximation is only valid for a large number of domains in each section. Here we calculate the PMF by explicitly considering the contribution from each domain. The nonlinearity profile $\chi_{\textsc{d}}(z)$ will consist of domains of nonlinear coefficients $\pm \chi^{(2)}$, with sign changes occurring at positions corresponding to the poling order and duty cycle (this is the case for the entire length of a typical periodically poled crystal). For example, in the section corresponding to $m{=}3$, where the duty cycle is $50\%$, the sign changes every $3\Lambda$, while for $m{=}6$, where the duty cycle is $\approx41.6\%$, the sign changes from ``$+$'' to ``$-$'' after $5\Lambda$ and back again after $7\Lambda$. The resulting PMF takes the form
\begin{eqnarray}
\Phi_{\textsc{d}}(\Delta k_{\mathrm{p}} )= \chi^{(2)}\sum_j s_j(\mathrm{e}^{{-}\ii\Delta k_{\mathrm{p}} z_j }-\mathrm{e}^{{-}\ii\Delta k_{\mathrm{p}} z_{j-1} })\,,
\end{eqnarray}
where $s_j$ is the sign of the $j$th domain and $z_{j-1}-z_{j}$ is the width of each domain. Figure~\ref{fig:comp_zoom} shows that as $\Delta k_{\mathrm{p}}$ departs from $0$, the two models begin to deviate. 

\begin{figure*}[t!]
  \begin{center}
   \includegraphics[width=0.8\textwidth]{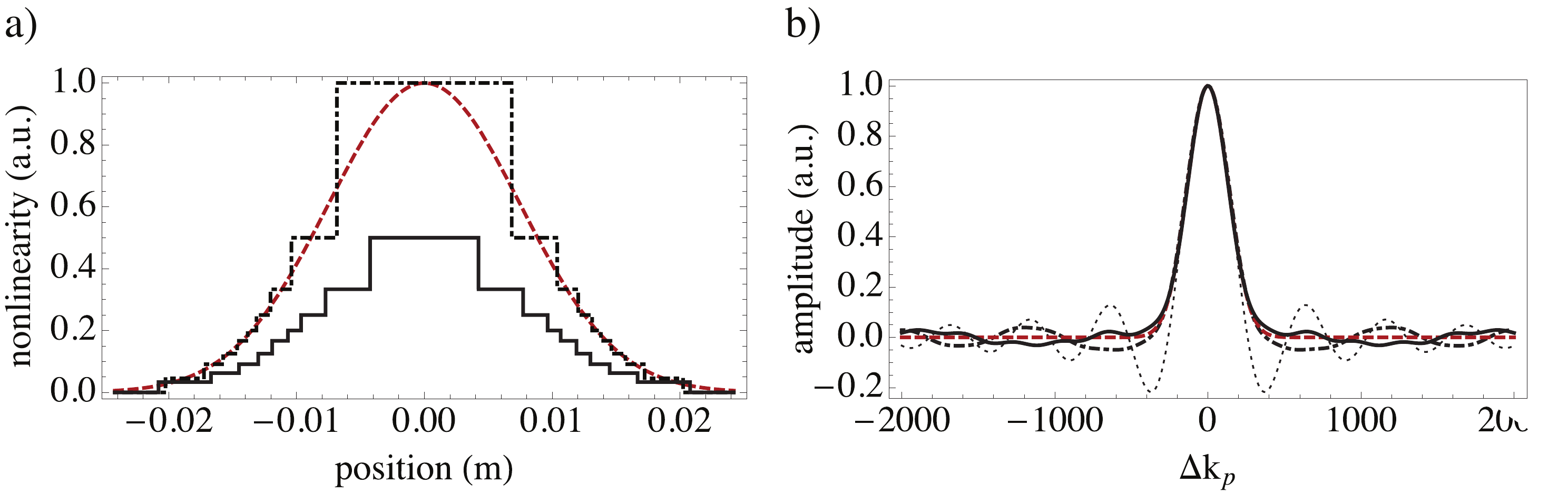}
    \vspace{-.5cm}
  \end{center}
    \caption{a) Nonlinearity profiles and b) corresponding PMFs for: $m=1$ tailored crystal (solid black line); $m=2$ tailored crystal (dot-dashed black line); and an ideal crystal with a Gaussian profile (dashed red line). The corresponding sinc PMF (thin dotted line) has been included for comparison. } 
  \label{fig:pure_photons}
\end{figure*}

However, as we discuss in the main text, there is very good agreement between the basic and detailed models in the region of interest, i.e. around $\Delta k_{\mathrm{p}}=0$.

\section{Crystal design for separable joint spectral amplitude}\label{sec:crystal_design}

Downconverted single-photons have strong spectral correlations which result in the degraded purity of a heralded state. A growing effort in engineering pulsed downconversion sources to produce spectrally decorrelated photons includes manipulating the crystal length, material, bandwidth and central frequency \cite{Kim2002, Walton2004, URen2006, Corona2007, URen2007, Kuzucu2008, Mosley2008, Garay-Palmett2007,Christ2009a} as well as filtering the pump field, prior to down-conversion, using an optical cavity \cite{Raymer2005}. Another method imposes group-velocity matching to limit these unwanted correlations \cite{Grice2001,URen2005}. However, the presence of side lobes---which arise from the sinc shape of the PMF---still calls for some level of filtering in order to achieve high-purity single photons. Spectral filtering is undesirable because it lowers the overall single-photon production rate as well as introducing photon-number mixedness which limits the allowable pump intensity \cite{Branczyk2009}. 

We show that, in combination with our method of modulating the crystal nonlinearity, group velocity matching can be used to create high-purity single photon states without the use of spectral filtering. Setting the relationship between the group velocities of the three interacting fields such that $k_p'=(k_s'+k_i')/2$ and picking the length of the crystal to be $L=\sqrt{8/\gamma \sigma_p^2(k_s'-k_i')^2}$ (where $\sigma_p$ is the pump width in s$^{-1}$ and $\gamma$ is defined below) generates a joint spectral amplitude (JSA), where both signal and idler modes have equal bandwidths. For a type-II ppKTP crystal, this corresponds to a crystal length $L{=}24.2$ mm and a periodicity of $\Lambda{=}68.4~\mu\mathrm{m}$, pumped with a $788$ nm laser with a 0.7nm FWHM which down converts to $1576$ nm in the signal and idler modes. However, for a standard crystal of constant nonlinearity, this will not result in completely pure states being generated, due to the side lobes in the sinc function. 

To eliminate the side lobes, we want to generate a Gaussian PMF,  $\Phi_{\mathrm{G}}(\omega_i,\omega_s){=}\exp(-\gamma(\Delta kL/2)^2)$, whose FWHM matches that of the PMF generated by a standard crystal, $\Phi(\omega_i,\omega_s){=}\mathrm{sinc}(\Delta k L/2)$. Substituting $\Phi_{\mathrm{G}}(\omega_i,\omega_s)$ as the PMF, we can now write the JSA as
\begin{equation}\nonumber
f(\omega_i,\omega_s)\propto\exp\Big({-}\frac{(\omega_i+\omega_s-2\mu)^2}{2\sigma_p^2}\Big)
\times\exp\Big({-} \frac{ \gamma \Delta k^2L^2}{4}\Big), \label{eq:JSA_sep}
\end{equation}
which, due to the rotational symmetry of the two-dimensional Gaussian function, is separable, i.e. $f(\omega_i,\omega_s)=g(\omega_i)h(\omega_s)$.

We numerically compare two designs for cpKTP crystals, Fig.~\ref{fig:pure_photons}, with a standard ppKTP crystal following \cite{Branczyk2009}. The results are summarised in Table I in the main text.

 \begin{figure}[b!]
  \begin{center}
   \includegraphics[width=.48\textwidth]{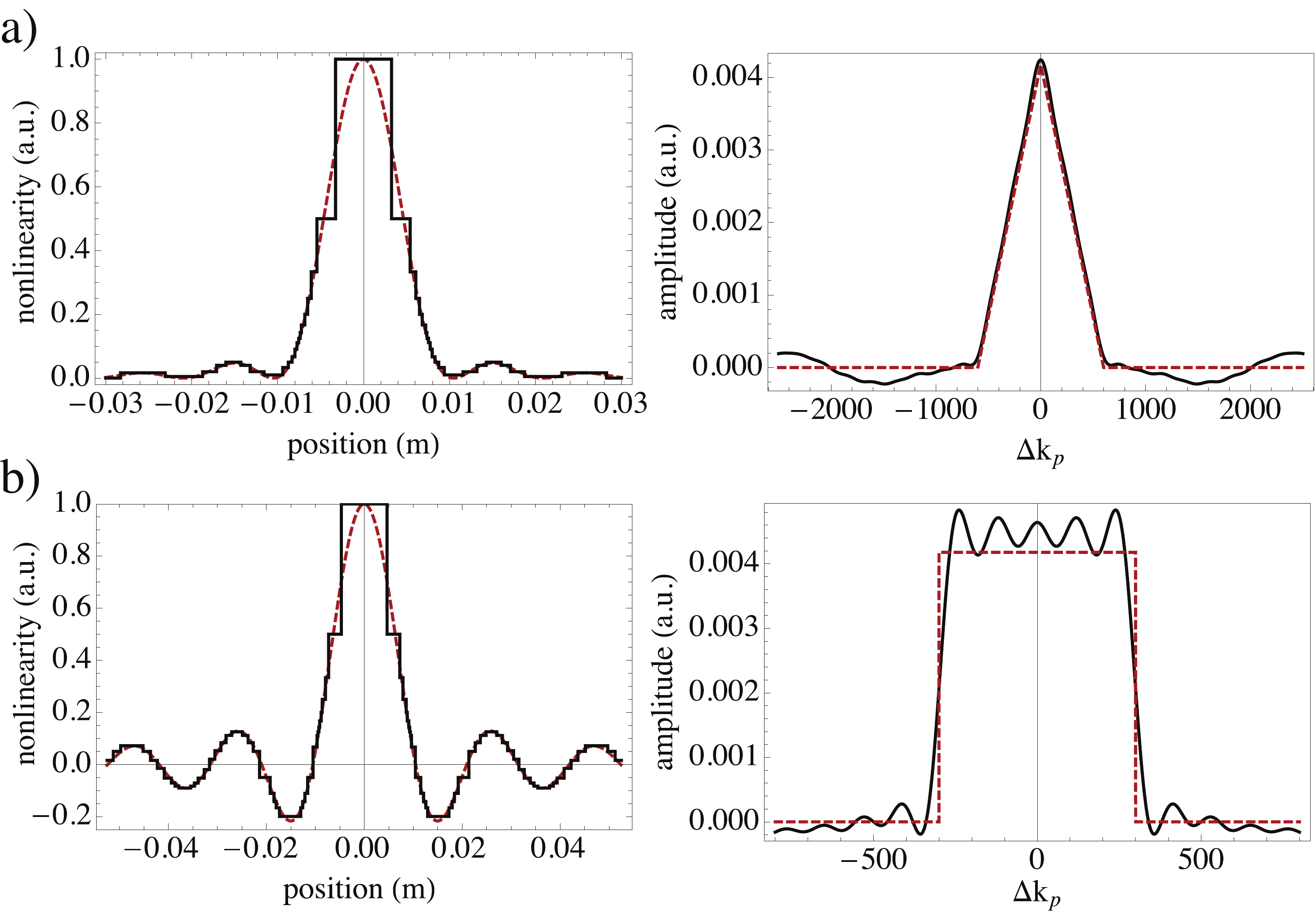}
  \end{center}
    \caption{a) Examples of tailored nonlinearity profiles, and corresponding phasematching functions, for a) a triangular phase-matching function and b) a top-hat function. The dashed red line shows the target functions and the black line the results of the discrete approximation.} 
  \label{fig:shapes}
\end{figure}

\section{Non-gaussian two-photon spectra}\label{sec:other_PMF}
The method introduced in this paper can be applied to the generation of almost arbitrarily shaped PMFs. As described above, the nonlinearity profile of the crystal should be tailored to the Fourier transform of the desired PMF. Figure \ref{fig:shapes} show examples of triangular and square shaped phase matching functions, as well as the required nonlinearity profile. Negative values of the nonlinearity, which are required to generate the square shape, can be implemented by inverting the relevant domain.

\end{document}